\documentclass[preprint,12pt,compress]{elsarticle}

\usepackage{amssymb}
\usepackage{amsmath}

\journal{Nuclear Physics B}

\usepackage{algorithmic}
\usepackage{algorithm}
\usepackage[caption=false,font=footnotesize,labelfont=rm,textfont=rm]{subfig}
\usepackage{booktabs} 
\usepackage{multirow}
\usepackage{pifont} 
\usepackage{cleveref}
\Crefname{figure}{Fig.}{Figs.}
\Crefformat{figure}{Fig.\thinspace#2#1#3}
\Crefformat{equation}{Eq.\thinspace#2#1#3}
\usepackage{microtype}
\newcommand{\proposed}{FedAPTA}

\begin{document}

\begin{frontmatter}




\title{FedAPTA: Federated Multi-task Learning for Heterogeneous Devices with Adaptive Layer-wise Pruning and Task-aware Aggregation}


\author[a,c]{Zhen~Yu} 
\author[a,c]{Yachao~Yuan\corref{corresponding}} 
\ead{chao910904@suda.edu.cn}
\author[a,c]{Jin~Wang} 
\author[a]{Zhipeng~Cheng} 
\author[b]{Jianhua~Hu} 

\cortext[corresponding]{Corresponding author: Yachao Yuan.}

\affiliation[a]{organization={School of Future Science and Engineering},
            addressline={Soochow University}, 
            city={Suzhou},
            state={Jiangsu},
            country={China}}
            
\affiliation[c]{organization={Key Laboratory of General Artificial Intelligence and Large Models in Provincial Universities (Soochow University)},
			city={Suzhou},
			state={Jiangsu},
			country={China}}
            
\affiliation[b]{organization={China Mobile (Suzhou) Software Technology Co., Ltd.},
            city={Suzhou},
            state={Jiangsu},
            country={China}}

\begin{abstract}
Federated Learning (FL) has shown considerable promise in Machine Learning (ML) across numerous devices for privacy protection, efficient data utilization, and dynamic collaboration. 
However, mobile devices typically have limited and heterogeneous computational capabilities, and different devices may even have different tasks. This client heterogeneity is a major bottleneck 
hindering the practical application of FL.
Existing work mainly focuses on mitigating FL's computation and communication overhead of a single task while overlooking the computing resource heterogeneity issue of different devices in FL. To tackle this, we design \proposed, a federated multi-task learning framework. FedAPTA overcomes computing resource heterogeneity through the developed layer-wise model pruning technique, which reduces local model size while considering both data and device heterogeneity. To aggregate structurally heterogeneous local models of different tasks, we introduce a heterogeneous model recovery strategy and a task-aware model aggregation method that enables the aggregation through infilling local model architecture with the shared global model and clustering local models according to their specific tasks. We deploy FedAPTA on a realistic FL platform and benchmark it against nine SOTA FL methods. The experimental outcomes demonstrate that the proposed \proposed~considerably outperforms the state-of-the-art FL methods by up to 4.23\%. Our code is available at https://github.com/Zhenzovo/FedAPTA.
\end{abstract}

%


\end{frontmatter}

\section{Introduction}
In the past ten years, the proliferation of billions of Internet of Things (IoT) devices has led to an unprecedented increase in data generation~\cite{khan2021federated_khaniot}. Federated Learning (FL)~\cite{mcmahan2017communicationfedavg} has emerged as a promising solution for enabling distributed model training across heterogeneous devices while preserving data privacy \cite{sun2022fedtar,guendouzi2023systematic,yuan2023computing_split,wei2024fedact,lin2024game_cpngame,wang2025novel_hete}. FL allows devices (i.e., participants of FL) to locally train models based on their confidential data and regularly transmit the model updates to the central server for aggregation, which facilitates data processing on distributed computing devices where data is generated. Additionally, FL distributes model training across devices to avoid the transmission of raw data to the central server, thereby reducing communication overhead and preserving user privacy \cite{mcmahan2017communicationfedavg,ezzeldin2023fairfed}.

Although FL has been effectively applied, it still faces serious challenges when handling multi-task scenarios with heterogeneous devices, particularly in terms of model deployment and aggregation. For model deployment, in most cases, heterogeneous devices participating in FL exhibit significant differences in computational capability, and many of them have limited resources. Therefore, if the same model is deployed on all devices without any adaptation, those with weaker computational capability will be unable to properly train the model or effectively participate in the FL process. This may hinder the progress of global model training and even cause the entire FL procedure to stall, ultimately affecting the normal training of the global model. For model aggregation, if the central server aggregates local models of all devices directly without task differentiation, the resulting global model will fail to accurately learn the task-specific knowledge associated with each device, which can cause performance reduction of the global model.

A number of pioneering studies have made significant efforts to tackle these two challenges. For the former, to enable devices with weaker computational capabilities to properly participate in the FL process, model architectures should be generated according to their computing resources. Some studies \cite{caldas2018expanding_feddrop,alam2022fedrolex,li2021hermes,jiang2022fedmp,jia2024fedlps,11052838} have explored model pruning strategies to address this issue. For example, FedMP~\cite{jiang2022fedmp} leverages a multi-armed bandit strategy to dynamically adjust the pruning ratio for each device, effectively adapting to heterogeneous devices while maintaining promising accuracy. For the same goal of accommodating the heterogeneous computational capabilities of devices, FedLPS~\cite{jia2024fedlps} proposes an adaptive channel-wise pruning method to generate lightweight local models. However, during the pruning process, existing research usually adopt a uniform pruning ratio for all layers to be pruned, without taking into account the varying importance levels of different layers, which can lead to poor performance of the pruned model. 
For the latter, to alleviate the issue of suboptimal global models caused by directly aggregating local models from different tasks, some studies cluster local models that belong to the same task before aggregation. For example, Clustered Federated Learning (CFL)~\cite{sattler2020clustered_cfl} clusters devices based on the similarity of their gradient updates, enabling the learning of more specialized models within each cluster.

\begin{figure*}
    \centering
    \includegraphics[width=\linewidth]{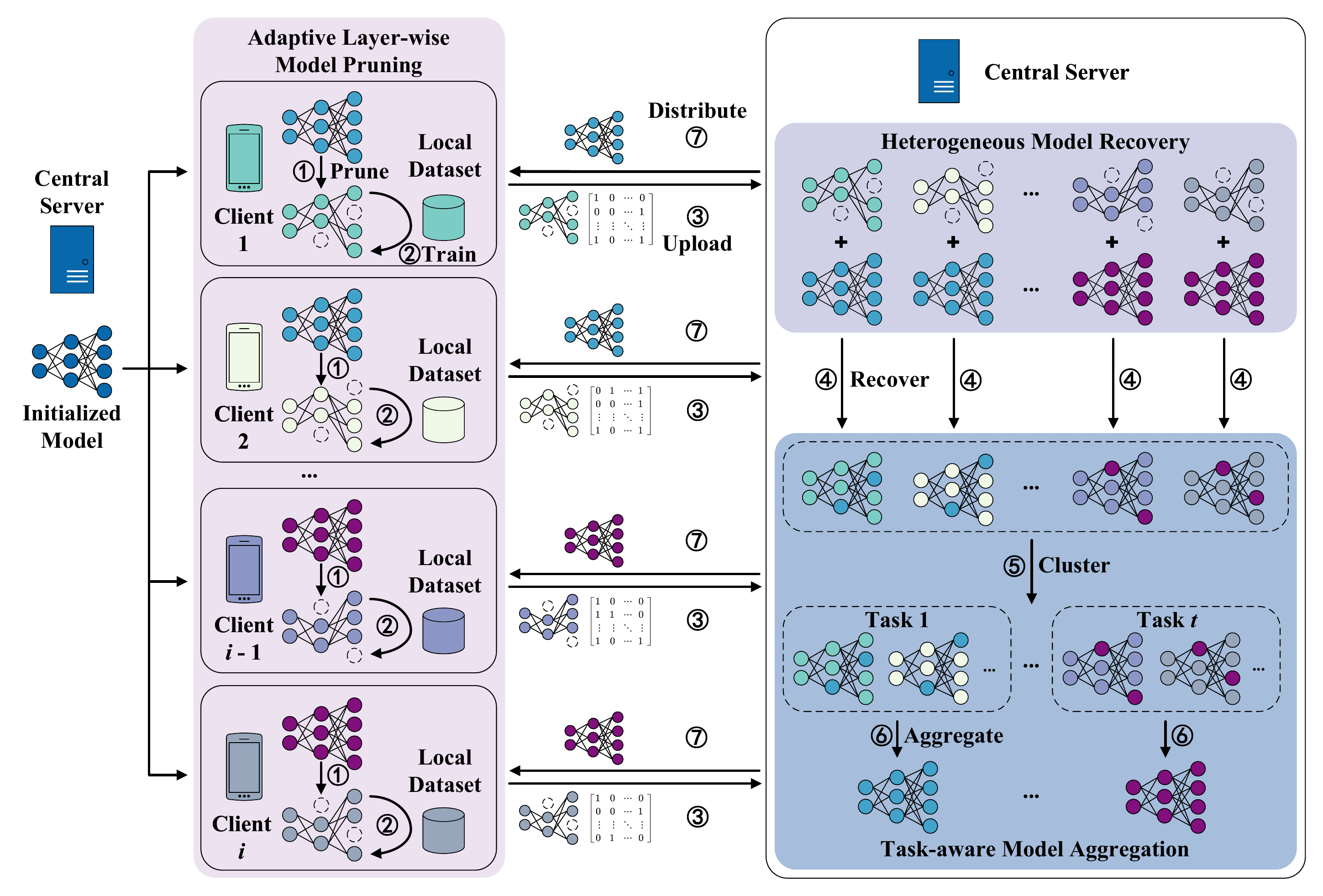}
    \caption{Overview of the proposed \proposed~framework. 
    To elaborate, \ding{192} After the global model is sent from the central server, the devices prune the global model by the adaptive layer-wise pruning to obtain the local model. \ding{193} The devices use its local data to train the local model. \ding{194} The devices upload the trained local model and the mask matrix to the central server. \ding{195} The central server utilizes global model information to recover the received local model. \ding{196} The central server clusters all local models using the distance matrix based on their updates. \ding{197} The central server aggregates the local models belonging to each task to obtain the corresponding global models. \ding{198} The central server distributes the global models belonging to each task to the corresponding devices.}
    \label{fig:framework}
\end{figure*}

In this paper, we propose a novel FEDerated multi-task learning framework for heterogeneous devices with Adaptive layer-wise Pruning and Task-aware Aggregation (FedAPTA) for addressing the model deployment and aggregation challenges in FL. Specifically, we allow each device to train a sub-model that is meticulously pruned to match its own local computing resources. 
Different from existing pruning methods \cite{caldas2018expanding_feddrop,alam2022fedrolex,li2021hermes}, we assign a unique pruning ratio to each layer rather than applying a uniform pruning ratio for all layers of a model to maintain satisfactory training performance even under relatively high pruning ratios. 
To aggregate the heterogeneous pruned models, we introduce a heterogeneous model recovery algorithm. Unlike existing algorithms \cite{li2021hermes,jiang2022fedmp} that only aggregate shared parameters across pruned local models, \proposed~leverages the information from the latest global model to assist in the reconstruction of a consistent architecture from the pruned models. It mitigates the issue wherein parameters pruned by other devices are no longer able to contribute to the learning process.
Additionally, since aggregating local models belonging to different tasks on the central server may lead to a suboptimal global model, we cluster devices that are performing the same task by their local models before aggregation.


Our main contributions are as follows:
 \begin{itemize}
    \item[$\bullet$] 
    We propose an adaptive layer-wise model pruning method for \proposed \quad that innovatively allocates pruning ratios based on layer importance. This reduces the parameter size of local models while maintaining accuracy, allowing devices to learn heterogeneous models tailored to their computational capabilities and ensuring the efficient progress of FL.
    \item[$\bullet$] 
    We design a heterogeneous model recovery algorithm for aggregating pruned heterogeneous local models. With the assistance of the existing global model information, \proposed~can effectively aggregate heterogeneous local models.
    \item[$\bullet$] We develop a task-aware aggregation strategy to address the performance degradation of global models caused by directly aggregating local models from different tasks. This method enables \proposed~to identify models associated with different tasks and selectively aggregate them, thereby preserving global model performance.
    \item[$\bullet$] We integrate the developed \proposed~into a realistic federated learning platform (FedML) and conduct comprehensive evaluations against nine SOTA frameworks. Experimental outcomes show that \proposed~maintains superior model accuracy compared with the aforementioned FL frameworks.
\end{itemize}

\begin{table}
\centering
\caption{List of key notations.}
\renewcommand{\arraystretch}{1.2}
\begin{tabular}{cl}
\toprule
\textbf{Notation} & \textbf{Description} \\
\midrule
$C$ & A set of devices \\
$i$ & The $i_{th}$ device in $C$ \\
$D_i$ & The local dataset of device $i$ \\
$\rho_i$ & The pruning ratio for the model of device $i$ \\
$R$ & Total FL rounds \\
$r$ & The $r_{th}$ FL round \\
$w$ & The initialized global model weights \\
$w_i$ & The model weights of device $i$ \\
$\hat{w_i}$ & The pruned model weights of device $i$ \\
$M_i$ & The pruning mask matrix of device $i$ \\
$\mathbb{R}^n$ & The n-dimensional real vector space \\
$P_i$ & A set of layer-wise pruning ratios for the model of device $i$ \\
$\rho_k$ & The pruning ratio for layer $k$ \\
$k$ & The $k_{th}$ trainable convolutional layer of a local model \\
$I$ & A set of importance scores for each candidate layer \\
$I_k$ & The importance score of layer $k$ \\
$N$ & A set of parameter numbers of each candidate layer \\
$N_k$ & The parameter count of layer $k$ \\
$d_{cos}$ & Cosine distance \\
$\Delta w_i$ & The local model updates of device $i$ \\
$T$ & A set of tasks \\
$t$ & The $t_{th}$ task in $T$ \\
$D_t$ & The set of local datasets on devices belonging to task $t$ \\
$w_t$ & The global model for task $t$ \\
\bottomrule
\end{tabular}
\label{tab:notations}
\end{table}

\section{Methodology}
    \subsection{Overview}
        In this study, we propose \proposed, a FL framework specifically designed to efficiently train task-specific models across heterogeneous devices, where different devices have different computing resources and may undertake different tasks. Unlike traditional FL frameworks, which focus on training and maintaining a shared model structure for all devices, \proposed~allows each device to possess a local model customized for its computational capability. The overall operational workflow of \proposed~is illustrated in \Cref{fig:framework}, which includes two fundamental steps: the training of local models on devices and the aggregation of models on the central server. Particularly, the central server uses a pre-trained model as the initialized global model $w$ and distributes it to all devices. $w$ can come from pre-trained models trained on openly accessible datasets. To facilitate efficient training in scenarios with heterogeneous client computational capabilities, we introduce an adaptive layer-wise model pruning method, which will be described in detail in Section~\ref{section:ap}. In addition, regarding to the model aggregation process, there are two central challenges that must be addressed: on one hand, to aggregate the pruned heterogeneous models, we utilize information from the latest global model to restore the complete structure of the local models, facilitating the aggregation of the pruned models (as described in Section~\ref{section:ag}); on the other hand, to aggregate different task-specific models, we cluster all devices by their local models to distinguish their tasks (as elaborated in Section~\ref{section:cl}). The complete implementation process of the proposed method is presented in~\Cref{alg:proposed}, and the key symbols used in this paper are listed in~\Cref{tab:notations}.
        
        We outline the steps of FL as follows: 1) The central server initializes the global model $w$ and distributes it to all devices. 2) Each device $i$ determines an appropriate pruning ratio \(\rho_i\) based on its individual computational capability, and applies the adaptive layer-wise model pruning technique to obtain a customized local model $\hat{w_i}$. 3) Each device $i$ trains its local model using its private dataset $D_i$, and subsequently uploads the pruned local model $\hat{w_i}$ and the mask matrix $M_i$ to the central server. 4) The central server restores the pruned parts of each local model $\hat{w_i}$ using the corresponding mask matrix $M_i$ and the latest global model $w_t$, where $t$ is the task executed by device $i$, thereby restoring local models to a consistent structural format suitable for aggregation. 5) The central server conducts clustering over the reconstructed local models in order to distinguish models based on their respective tasks. 6) The central server aggregates models within each identified cluster, thereby aggregating local models related to the same task $t$. 7) The central server distributes the aggregated task-specific global models $w_t$ to the corresponding devices for further training.

\begin{algorithm}[t]
    \renewcommand{\algorithmicrequire}{\textbf{Input:}}
    \renewcommand{\algorithmicensure}{\textbf{Output:}}
    \caption{FedAPTA.}
    \label{alg:proposed}
    \begin{algorithmic}[1] 
        \REQUIRE Pruning ratios $\rho_i$, local datasets $D_i$($i \in C$).
        \ENSURE Global models $\{w_t \mid t \in T\}$.
        
        \STATE The central server initializes the global model $w$;
        
        \FOR{Each round $r \in \{1, 2, \cdots, R\}$}
            \IF{$r==1$}
                \STATE Central server sends $w$ to devices;
            \ENDIF
            \FOR{Each device $i \in C$}
                \STATE device $i$ prunes its local model $w_i$ based on the pruning ratio $\rho_i$ using~\Cref{alg:pr}, obtaining the pruned local model $\hat{w_i}$;
                \STATE device $i$ uses its local data $D_i$ to train its pruned local model $\hat{w_i}$;
                \STATE device $i$ uploads the trained local model $\hat{w_i}$ and the mask matrix $M_i$ to the central server;
            \ENDFOR
            \STATE The central server utilizes~\Cref{eq:rec} with the assistance of the corresponding global model and the mask matrix $M_i$ to recover the pruned local model $\hat{w_i}$;
            \STATE The central server clusters all local models using~\Cref{alg:clu} to identify the task $t \in T$ corresponding to each model;
            \STATE The central server aggregates the local models that belong to the same task $t$ and achieves a global model $w_t$ for each task;
            \STATE The central server distributes $w_t$ to the devices that are performing the task $t$;
        \ENDFOR

        \RETURN Global models $\{w_t \mid t \in T\}$.
    \end{algorithmic}
\end{algorithm}

\subsection{Adaptive Layer-wise Model Pruning}
\label{section:ap}
Existing FL frameworks, such as FedAvg~\cite{mcmahan2017communicationfedavg}, commonly adhere to a relatively fixed scheme: devices are assigned local models that share the same architecture as the global model, while local training and updates are conducted on this unified model architecture. While this design simplifies the model aggregation and synchronization processes, it exhibits significant limitations in practice. Specifically, since devices in FL have different computational capabilities, enforcing the optimization of an identical model structure across all devices often leads to inefficiency, thereby restricting the applicability of FL in heterogeneous devices environments.
        
To address these challenges, we propose \proposed. Distinct from existing FL frameworks that require all devices to train an identical model, in \proposed, after the global model is sent from the central server, each device \(i \in C\) prunes the global model using the proposed adaptive layer-wise model pruning method in order to derive a local model that is compatible with its computational capability. In this context, the pruning ratio \(\rho_i\) is determined by device \(i\), based on an assessment of its computing resources. 
Through this adaptive layer-wise pruning process, each device is capable of learning a local model that is structurally sparse. 
This pruning strategy enables heterogeneous devices to train models that are tailored to their computational capabilities, thereby avoiding the impact on the efficiency of FL.
        
During the FL training process, after receiving the global model, each device $i$ performs a pruning operation on the model using a pruning ratio \(\rho_i\). In contrast to approaches such as FedDrop~\cite{caldas2018expanding_feddrop}, which adopts a uniform pruning ratio across all participating devices in FL, \proposed~allows each device to decide its pruning ratio \(\rho_i\) based on its unique computational capability, improving the flexibility of FL. 
What's more, for each trainable convolutional layer of the local model, \proposed~applies a differentiated pruning ratio that is adaptively assigned according to the assessed importance of each candidate layer to better preserve critical features and improve the balance between pruning effectiveness and model performance. 

To begin with, for the model deployed on each device $i$, \proposed~evaluates the importance score of each candidate layer $k$ within the model $w_i$ by employing the L1 norm~\cite{liu2017learning_pr1}, obtaining the importance score vector $I=[I_1, \cdots, I_k, \cdots, I_n]$. Subsequently, while ensuring that the overall pruning ratio of the model remains consistent with the predefined whole pruning ratio target $\rho_i$, the pruning ratio $\rho_k$ for each candidate layer $k$ is assigned based on its parameter quantity and importance score. Specifically, let the parameter count of each candidate layer $k$ be represented by the vector $N = [N_1, \cdots, N_k, \cdots, N_n]$; this vector is then reordered according to the descending order of $I$. The objective is to assign a pruning ratio $\rho_k$ to each candidate layer $k$, where $\rho_k$ is between 0 and 1, such that the aggregate pruning across all layers adheres to the specified overall pruning ratio $\rho_i$, while the sequence of assigned pruning ratios exhibits a non-decreasing structure, it means that the layers that are deemed less important are assigned a pruning ratio that is no less than the ratios of more important layers. Let $P_i=[\rho_1, \cdots, \rho_k, \cdots, \rho_n]$ represent the optimization variable, then the formulation of the optimization problem is given as follows:
\begin{equation}
\begin{aligned}
\min_{P_i \in \mathbb{R}^n} \quad & 0, \\
\text{s.t.} \quad
& \sum_{k=1}^{n} N_{k} \times \rho_{k} = \rho_i \times \sum_{k=1}^{n} N_{k}, \\
& \rho_{k+1} - \rho_{k} \ge 0, \quad \forall\, k = 1, \dots, n-1, \\
& 0 \le \rho_{k} \le 1, \quad \forall\, k = 1, \dots, n,
\end{aligned}
\label{eq:cal_pr}
\end{equation}
where $\mathbb{R}^n$ denotes the n-dimensional real vector space. Once the layer-level pruning ratios are determined, each trainable convolutional layer is pruned individually based on its L1 importance to achieve the whole model pruning ratio $\rho_i$. Formally, the pruning step is represented as follows:
\begin{equation}
    \hat{w_i}=w_i \odot M_i,
\end{equation}
where $\odot$ denotes element-wise multiplication, $w_i$ represents parameters of the model of device $i$, and $M_i$ is a binary mask matrix employed to identify the channels to be pruned. In $M_i$, an element value of 0 denotes the corresponding channel will be pruned, while an element value of 1 denotes the corresponding channel will be retained. The overall process of the proposed adaptive layer-wise model pruning method is presented in~\Cref{alg:pr}.

This method enables heterogeneity and sparsity of model structures, effectively adapting to the different computational capabilities of devices, thereby enhancing the practical applicability of FL.

\begin{algorithm}[t]
    \renewcommand{\algorithmicrequire}{\textbf{Input:}}
    \renewcommand{\algorithmicensure}{\textbf{Output:}}
    \caption{Adaptive Layer-wise Model Pruning.}
    \label{alg:pr}
    \begin{algorithmic}[1]
        \REQUIRE Local model $w_i$, pruning ratio $\rho_i$.
        \ENSURE Pruned local model $\hat{w_i}$.
        \FOR{Each candidate layer $k$ in $w_i$}
            \STATE Count the number of parameters $N_k$ in $k$;
            \STATE Calculate the L1 importance $I_k$ of $k$;
        \ENDFOR
        \STATE Sort layer indices by $I_k$ in descending order, get permutation $\pi$;
        \STATE Reorder $I$ and $N$ according to $\pi$;
        \STATE Solve the constrained optimization problem described by~\Cref{eq:cal_pr}, obtaining the pruning ratio $\rho_k$ for each candidate layer $k$;
        \STATE Prune each candidate layer $k$ with the pruning ratio $\rho_k$;
        \RETURN Pruned local model $\hat{w_i}$.
    \end{algorithmic}
\end{algorithm}

    \subsection{Heterogeneous Model Recovery}
    \label{section:ag}
        Adaptive layer-wise model pruning offers a significant advantage in FL, particularly when applied across devices with different computing resources. By systematically removing less important parameters in a model on a per-layer basis, this method effectively reduces the overall size of the local models. Consequently, it significantly reduces computational cost, which is especially beneficial for heterogeneous devices. This approach allows each device to determine the pruning ratio for its local model according to its computational capability, enabling more flexible and on-demand pruning. By enabling heterogeneous devices to participate in FL using models with different sizes, it avoids a decrease in the efficiency of FL caused by deploying a uniform model architecture on devices with heterogeneous computing resources.
        
        However, despite these benefits, the proposed adaptive layer-wise pruning technique inherently includes structural heterogeneity among devices' models. Specifically, since each device independently prunes its local model according to its computational capability, the resulting model architectures can vary significantly from one device to another. It presents a major challenge for model aggregation, which is a fundamental operation in FL, thereby making models incompatible with popular and widely adopted model aggregation algorithms, such as FedAvg~\cite{mcmahan2017communicationfedavg}, as mismatched architectures prevent straightforward averaging or fusion. As a result, application of such algorithms to models with heterogeneous structures may result in degraded global model performance or even complete aggregation failure. To cope with these structural discrepancies, several heterogeneous model aggregation approaches have been proposed recently. For example, FedMP~\cite{jiang2022fedmp} focuses on aggregating the parameters that are shared among heterogeneous local models. Nevertheless, if certain parameters of a device's local model have been pruned on other devices, the device will be unable to benefit from those parameters. 
        
        To overcome this challenge, we propose a heterogeneous model recovery method to support the aggregation of structurally heterogeneous models in FL. The key idea of our method is to recover the pruned parameters in each device’s model using the latest global model, thereby creating a full-structure model that can be safely and meaningfully aggregated by the central server. This strategy ensures that all parameters are taken into account during the aggregation process, whether retained or pruned by devices. As a result, it enables each device to fully participate in the collaborative learning process of FL, regardless of its specific pruning decisions, thus maximizing cross-device knowledge sharing. Specifically, in each round of FL, after receiving the trained and pruned local models from each device, the central server first uses the latest global model and the mask matrix \( M_i \) of the device \( i \) to recover the pruned parameters in the pruned local model $\hat{w_i}$. The recovery operation can be represented as follows:
        \begin{equation}
            w_i=\hat{w_i}+w_t \odot (1-M_i),
            \label{eq:rec}
        \end{equation}   
        where $\hat{w_i}$ is the model to be recovered and $w_t$ is the latest global model of task $t$, which is executed by device $i$. The term $(1-M_i)$ effectively identifies the pruned parameters of local models, then the central server replaces these pruned values with their corresponding entries from the latest global model. 
        
        This approach enables the central server to reconstruct a complete version of the model for each device by recovering the pruned parameters of the heterogeneous local models. It ensures that even models with different structures can benefit from each other, significantly improving the robustness and performance of FL in heterogeneous environments.

\subsection{Task-aware Model Aggregation}
\label{section:cl}
In this section, we implement a task-aware aggregation method by device clustering, which is specifically designed to ensure the performance of the global model for multi-task scenarios. In traditional FL settings, it is generally assumed that all devices are working on the same global task and hence can freely share updates for direct aggregation. However, in real-world scenarios, participating devices are usually engaged in different tasks. This type of task heterogeneity presents a significant challenge for conventional aggregation methods, as directly combining models from different tasks may yield suboptimal or even misleading global models. Such models may fail to serve any individual task effectively, and in some cases, may even result in harmful interference across tasks. To solve this issue, the primary objective of our proposed method is to distinguish models that belong to different tasks. By doing so, the central server can perform task-aware aggregations, thereby preserving task-specific knowledge and avoiding negative transfer between unrelated tasks. It is particularly beneficial in multi-task FL scenarios. 

As aforementioned, for traditional FL frameworks like~\cite{mcmahan2017communicationfedavg}, which involves two fundamental processes: 1) The aggregation and broadcasting of models on the server side; 2) The training and uploading of models on the device side. Our modifications predominantly focus on the server side, while leaving the device-side operations unchanged. This design ensures that our method can be easily integrated into current FL frameworks. 
Specifically, before the model aggregation step, 
we apply a model clustering step to group devices, thereby identifying which devices can collaboratively work on the same task.

To achieve better performance in practice, we use the cosine similarity of the local model updates to compute the distance \(d_{cos}\) as the clustering criterion. Specifically, the distance is expressed as follows:
\begin{equation}
    d_{cos}=1-\cos(\Delta w_i,\Delta w_j),
    \label{eq:dis}
\end{equation}
where $\Delta w_i=w_i-w_t$, and \(\Delta w_i, \Delta w_j\) are the local model updates on the device \(i, j\) respectively. A small \(d_{cos}\) value indicates that two devices share similar update directions and suggests a high probability of working on the same task, while a larger value indicates dissimilar updates and hence possible task divergence. Additionally, computing this distance across all model parameters can be computationally expensive and may include noisy signals from layers unrelated to task-specific features. According to pFedGraph~\cite{ye2023personalized_pfedgraph}, the last few fully connected layers are sufficient to reflect the fine-grained similarity of models, so we use the cosine distance of the parameters' update values in the last few fully connected layers to approximate \(d_{cos}\) to reduce computational cost. 
This approximation achieves a balance between clustering efficiency and performance by filtering out irrelevant noise from early layers. 

After computing the cosine distances \(d_{cos}\) between each pair of the received local models, a square matrix (i.e., distance matrix) is constructed where each entry represents the distance between a pair of models.
Then the central server applies HDBSCAN~\cite{campello2013density_hdbscan} based on this matrix to identify groups of devices associated with the same task. 
Unlike traditional clustering algorithms, HDBSCAN operates without the number of clusters to be specified in advance, which aligns well with FL, where the number of distinct tasks may vary and is not known as a priori. Moreover, HDBSCAN is resilient to noise and outliers, making it particularly suitable for federated settings where the data may be uneven and noisy. Each resulting cluster ideally corresponds to a subset of devices working on the same task \(t \in T\). 
The clustering process is described in~\Cref{alg:clu}.

Subsequently, based on the clustering results, \proposed~aggregates models for each task \(t\) individually using a weighted average scheme. The weight assigned to each device is based on its local dataset size, so that devices with larger dataset sizes play a more prominent role in updating the global model. The aggregation operation is formally defined as:
\begin{equation}
    w_{t}=\sum_{i \in C_t}\frac{|D_{i}|}{|D_t|}w_i,
\label{eq:agg}
\end{equation}   
where $C_t$ denotes the set of devices belonging to task $t$, $D_i$ denotes the local dataset on device $i$, and $D_t$ denotes the set of local datasets of all devices belonging to task $t$. For each task \(t\), the global model $w_{t}$ is computed by aggregating the recovered local models according to \Cref{eq:agg} until the aggregation process is complete. 
Finally, the central server distributes the updated global model $w_{t}$ to all devices associated with task \(t\). These devices will use this global model as the initialization for their next round. 

This approach ensures that devices only learn from devices with the same goal, thereby maximizing positive knowledge transfer and minimizing negative interference. It is beneficial in multi-task FL, where institutions or user groups may work on different tasks. By clustering, our method supports task specialization while preserving the overall benefits of model sharing in FL settings.

\begin{algorithm}[t]
    \renewcommand{\algorithmicrequire}{\textbf{Input:}}
    \renewcommand{\algorithmicensure}{\textbf{Output:}}
    \caption{Device Clustering based on local models.}
    \label{alg:clu}
    \begin{algorithmic}[1]
        \REQUIRE Local models $w_i(i \in C)$, global model $w$.
        \ENSURE All tasks $\{t \mid t \in T\}$.
        \FOR{Each device pair $(i, j) \in C$}
            \STATE Compute model updates of $w_i$ and $w_j$, obtain $\Delta w_i$ and $\Delta w_j$;
            \STATE Compute cosine distance \(d_{cos}\) between $\Delta w_i$ and $\Delta w_j$ using~\Cref{eq:dis};
        \ENDFOR
        \STATE Construct distance matrix;
        \STATE Apply HDBSCAN to obtain the task $t \in T$ corresponding to each model;
        \RETURN $\{t \mid t \in T\}$.
    \end{algorithmic}
\end{algorithm}

\section{Experimental Evaluation}
In this section, we implement the proposed \proposed~framework on a real-world FL platform, FedML~\cite{he2020fedml}, followed by a performance comparison with nine existing frameworks. Then we investigate how varying pruning ratios affect the learning performance of \proposed. Finally, we evaluate the effectiveness of various model similarity metrics in distinguishing models belonging to different tasks.

\subsection{Experimental Setting}
\subsubsection{FL environment}
We simulate fifty heterogeneous devices and set up five distinct classification tasks, with each task being assigned to a group of ten devices. We employ ResNet18~\cite{he2016deepresnet} and ShuffleNetV2~\cite{zhang2018shufflenet} models to carry out these classification tasks. 
Following the setup of FedLPS~\cite{jia2024fedlps}, we use the pretrained model on ImageNet~\cite{russakovsky2015imagenet} as the initialized global model and freeze the first quarter of the layers during training. 
Moreover, inspired by FedDC~\cite{gao2022feddc}, we replace the Batch Normalization layers in the model with Group Normalization~\cite{wu2018group} layers, and learn a local drift variable to track and correct the gap between the local model and the global model, in order to mitigate the effect of the local data distributions on model performance. 
In our simulation experiments, 
the FL training process includes all heterogeneous devices. Experimental evaluations are performed on a GPU server equipped with four NVIDIA RTX 3090 GPUs.

\begin{table}
\centering
\caption{Hyper-parameters used in FL training.}
\renewcommand{\arraystretch}{1.2}
\begin{tabular}{lc}
\toprule
\textbf{Hyper-parameter} & \textbf{Value} \\
\midrule
Learning rate & 0.1 \\
Learning decay & 0.998 \\
Weight decay & 0.001 \\
Local epoch & 5 \\
Global round & 100 \\
Number of devices & 10 \\
Parameter $\alpha$ of LDA & 0.5 \\
\bottomrule
\end{tabular}
\label{tab:hyper-params}
\end{table}

\subsubsection{Datasets and data partition}
We explore on five widely recognized datasets: MNIST~\cite{lecun1998gradientmnist}, FashionMNIST~\cite{xiao2017fashionmnist}, SVHN~\cite{netzer2011readingsvhn}, CIFAR10~\cite{krizhevsky2009learningcifar10}, and EMNIST~\cite{cohen2017emnist} to simulate the classification tasks. Under the independent and identically distributed (i.i.d.) setting, each dataset was evenly allocated across the devices. In the non-independent and identically distributed (non-i.i.d.) setting, following~\cite{jia2024fedlps,luo2021noDirichlet}, we employ the Latent Dirichlet Allocation (LDA) method to construct the non-i.i.d. data. Within LDA, we use the conventional setting of $\boldsymbol{\alpha} = 0.5$ to simulate non-i.i.d. data distributions, where $\alpha$ is utilized to regulate the level of data heterogeneity.

\subsubsection{Comparison frameworks}
We compare the proposed \proposed~framework with FedAvg~\cite{mcmahan2017communicationfedavg}, Ditto~\cite{li2021ditto}, FedProx~\cite{li2020federatedfedprox}, FedGen~\cite{zhu2021datafedgen}, MOON~\cite{li2021modelmoon}, FedBABU~\cite{oh2022fedbabu}, FedNTD~\cite{lee2022preservation_fedntd}, FedLC \quad \cite{zhang2022federated_fedlc} and FedLPS~\cite{jia2024fedlps}. FedAvg is a classical FL framework that requires each client to update the entire model, thus the training efficiency of FL with heterogeneous devices is affected by participants with lower computational capabilities. Ditto enables devices to maintain personalized models, thereby adapting effectively to the local data distributions. 
FedProx introduces a regularization term to encourage local models to stay close to the global model from the same task, while allowing devices with limited computational capabilities to perform fewer local updates. FedGen addresses data heterogeneity issues in FL by training a generator on the server side that aggregates knowledge from devices as prior knowledge for local training of the same task. MOON employs contrastive learning on model representations, utilizing the discrepancies between the current local model and the global model from the same task to guide local training, effectively mitigating challenges arising from non-i.i.d. data distributions. FedBABU incorporates attention-based representation learning to enhance the model’s representational power, thereby improving performance in image classification tasks. FedNTD mitigates the forgetting problem of the global model during the FL process by preserving non-ground-truth class information during local training, which leads to improved performance on non-i.i.d. data. FedLC calibrates logits to reduce model overfitting on minority classes within a task, enhancing classification accuracy under long-tail or imbalanced data distributions. 
FedLPS reduces resource consumption for multi-task learning in heterogeneous FL through local parameter sharing and a channel-wise model pruning algorithm. 
To ensure reliability, experiments are repeated three times to derive average results. Moreover, to ensure a just comparison, the same initialized global model was adopted for both \proposed~and all comparative frameworks during the experiments. Detailed information regarding the training hyperparameters is provided in \Cref{tab:hyper-params}.

\begin{table*}
\centering
\caption{The ResNet18 model accuracy (\%) comparison under both i.i.d. and non-i.i.d. settings.}
\label{cmp:r18}
\renewcommand{\arraystretch}{1.2}
\resizebox{\linewidth}{!}
{
\begin{tabular}{cccccccc}
\toprule
\textbf{Data} 
& \multirow{2}{*}{\textbf{FL frameworks}}
& \multirow{2}{*}{\textbf{MNIST}}
& \textbf{Fashion-} 
& \multirow{2}{*}{\textbf{SVHN}}
& \multirow{2}{*}{\textbf{CIFAR10}}
& \multirow{2}{*}{\textbf{EMNIST}}
& \multirow{2}{*}{\textbf{Average}} \\
\textbf{partition} & & & 
\textbf{MNIST} & & & & \\
\midrule
\multirow{12}{*}{\shortstack{i.i.d.\\partition}} 
& FedAvg~\cite{mcmahan2017communicationfedavg}    & 92.85 & 87.72 & 79.04 & 70.79 & 79.29 & 81.94 \\
& Ditto~\cite{li2021ditto}                  & 93.06 & 87.73 & 79.21 & 70.66 & 79.30 & 82.02 \\
& FedProx~\cite{li2020federatedfedprox}            & 93.01 & 87.69 & 79.29 & 70.52 & 79.26 & 81.96 \\
& FedGen~\cite{zhu2021datafedgen}                 & 92.94 & 87.86 & 79.27 & 70.37 & 79.18 & 81.93 \\
& MOON~\cite{li2021modelmoon}                   & 92.93 & 87.63 & 79.32 & 70.43 & 79.32 & 81.93 \\
& FedBABU~\cite{oh2022fedbabu}              & 92.88 & 87.54 & 79.25 & 70.45 & 79.37 & 81.90 \\
& FedNTD~\cite{lee2022preservation_fedntd}        	& 93.07 & 87.55 & 79.16 & 70.66 & 79.21 & 81.93 \\
& FedLC~\cite{zhang2022federated_fedlc}         	& 92.97 & 87.63 & 79.21 & 70.66 & 79.29 & 81.95 \\
& FedLPS~\cite{jia2024fedlps}    &  \underline{98.06} & \underline{89.18} & \underline{91.64} & \underline{80.10} & \underline{85.01} & \underline{88.80} \\
[0.3ex]\cline{2-8}\noalign{\vskip 0.5ex}
& \textbf{\proposed~(Ours)} & \textbf{98.94} & \textbf{91.16} & \textbf{92.17} & \textbf{80.40} & \textbf{85.04} & \textbf{89.54} \\
\midrule
\multirow{12}{*}{\shortstack{Non-i.i.d.\\partition}} 
& FedAvg~\cite{mcmahan2017communicationfedavg}    & 91.73 & 86.46 & 76.02 & 65.35 & 77.2 & 79.35 \\
& Ditto~\cite{li2021ditto}                  & 91.82 & 86.38 & 75.89 & 65.98 & 77.19 & 79.45 \\
& FedProx~\cite{li2020federatedfedprox}            & 91.73 & 86.44 & 76.04 & 65.82 & 77.2 & 79.45 \\
& FedGen~\cite{zhu2021datafedgen}                 & 91.9 & 86.43 & 75.99 & 65.95 & 77.03 & 79.46 \\
& MOON~\cite{li2021modelmoon}                   & 91.75 & 86.5 & 75.92 & 65.64 & 77.23 & 79.4 \\
& FedBABU~\cite{oh2022fedbabu}              & 91.82 & \underline{86.62} & 76.15 & 66.35 & 77.5 & 79.69 \\
& FedNTD~\cite{lee2022preservation_fedntd}        	& 91.65 & 86.5 & 76.14 & 65.55 & 77.18 & 79.4 \\
& FedLC~\cite{zhang2022federated_fedlc}         	& 91.76 & 86.26 & 76 & 65.81 & 77.09 & 79.38 \\
& FedLPS~\cite{jia2024fedlps}    &  \underline{97.25} & 84.24 & \underline{88.22} & \underline{67.49} & \underline{79.65} & \underline{83.37} \\
[0.3ex]\cline{2-8}\noalign{\vskip 0.5ex}
& \textbf{\proposed~(Ours)} & \textbf{98.75} & \textbf{89.13} & \textbf{89.82} & \textbf{68.22} & \textbf{83.44} & \textbf{85.87} \\
\bottomrule
\end{tabular}
}
\end{table*}

\subsection{Comparison of the Model Accuracy}
We compare the model accuracy of \proposed~with FedAvg, Ditto, FedProx, FedGen, MOON, FedBABU, FedNTD, FedLC, and FedLPS under both i.i.d. and non-i.i.d. settings. In this experiment, pruning ratios $\rho$ of $\{0, 0.2, 0.4, 0.6, 0.8\}$ were assigned to every group of five devices. 
\Cref{cmp:r18} presents the accuracy results of the ResNet18 model on i.i.d. and non-i.i.d. datasets, while \Cref{cmp:sfnet} details the accuracy outcomes for the ShuffleNetV2 model. The superiority of \proposed~can be attributed to three key factors: 
first, the models tailored for devices undergo meticulous pruning to preserve their performance as much as possible. 
Second, the heterogeneous model recovery algorithm uses information contained within the global model to assist in aggregating local models, facilitating improved knowledge transfer, and enabling models to learn more effectively from other devices. 
Third, the central server computes the cosine similarity among all received local models and employs HDBSCAN based on these similarities to cluster local models belonging to different tasks separately, thereby guaranteeing that the aggregated global models can correctly learn from knowledge based on their assigned tasks.

\begin{table*}
\centering
\caption{The ShuffleNetV2 model accuracy (\%) comparison under both i.i.d. and non-i.i.d. settings.}
\label{cmp:sfnet}
\renewcommand{\arraystretch}{1.2}
\resizebox{\linewidth}{!}
{
\begin{tabular}{cccccccc}
\toprule
\textbf{Data} 
& \multirow{2}{*}{\textbf{FL frameworks}}
& \multirow{2}{*}{\textbf{MNIST}}
& \textbf{Fashion-} 
& \multirow{2}{*}{\textbf{SVHN}}
& \multirow{2}{*}{\textbf{CIFAR10}}
& \multirow{2}{*}{\textbf{EMNIST}}
& \multirow{2}{*}{\textbf{Average}} \\
\textbf{partition} & & & 
\textbf{MNIST} & & & & \\
\midrule
\multirow{12}{*}{\shortstack{i.i.d.\\partition}} 
& FedAvg~\cite{mcmahan2017communicationfedavg}    & 88.21 & 82.62 & 66.76 & 58.45 & 75.48 & 74.30 \\
& Ditto~\cite{li2021ditto}                  & 88.07 & 82.61 & 67.38 & 57.99 & 75.39 & 74.29 \\
& FedProx~\cite{li2020federatedfedprox}            & \underline{88.37} & 82.59 & 67.64 & 58.82 & 75.38 & 74.56 \\
& FedGen~\cite{zhu2021datafedgen}                 & 88.15 & 82.65 & 67.25 & 58.15 & 75.34 & 74.31 \\
& MOON~\cite{li2021modelmoon}                   & 88.27 & 82.61 & 67.80 & 57.91 & 75.50 & 74.42 \\
& FedBABU~\cite{oh2022fedbabu}              & 88.16 & \underline{82.89} & 67.89 & 57.90 & 75.75 & 74.52 \\
& FedNTD~\cite{lee2022preservation_fedntd}        	& 87.74 & 82.71 & 67.53 & 57.63 & 75.47 & 74.22 \\
& FedLC~\cite{zhang2022federated_fedlc}         	& 88.14 & 82.68 & 67.60 & 57.76 & 75.43 & 74.32 \\
& FedLPS~\cite{jia2024fedlps}    &  84.57 & 79.86 & \underline{71.99} & \textbf{66.03} & \textbf{83.50} & \underline{77.19} \\
[0.3ex]\cline{2-8}\noalign{\vskip 0.5ex}
& \textbf{\proposed~(Ours)} & \textbf{94.84} & \textbf{86.86} & \textbf{73.20} & \underline{62.04} & \underline{82.87} & \textbf{79.96} \\
\midrule
\multirow{12}{*}{\shortstack{Non-i.i.d.\\partition}} 
& FedAvg~\cite{mcmahan2017communicationfedavg}    & 85.13 & 80.25 & 63.06 & 52.57 & 72.65 & 70.73 \\
& Ditto~\cite{li2021ditto}                  & 85.44 & 80.06 & 63.18 & 53.49 & 72.76 & 70.99 \\
& FedProx~\cite{li2020federatedfedprox}            & \underline{85.84} & 79.92 & 63.20 & 52.58 & 72.80 & 70.87 \\
& FedGen~\cite{zhu2021datafedgen}                 & 85.40 & 79.91 & 62.93 & 52.56 & 72.83 & 70.72 \\
& MOON~\cite{li2021modelmoon}                   & 85.82 & 79.86 & 63.24 & 52.40 & 72.78 & 70.82 \\
& FedBABU~\cite{oh2022fedbabu}              & 85.35 & \underline{80.34} & 62.92 & 53.54 & 73.21 & \underline{71.07} \\
& FedNTD~\cite{lee2022preservation_fedntd}        	& 85.48 & 80.07 & 62.98 & 52.13 & 72.76 & 70.69 \\
& FedLC~\cite{zhang2022federated_fedlc}         	& 85.31 & 80.31 & 63.62 & 52.61 & 72.81 & 70.93 \\
& FedLPS~\cite{jia2024fedlps}    &  78.49 & 74.07 & \underline{65.09} & \underline{59.71} & \underline{77.55} & 70.98 \\
[0.3ex]\cline{2-8}\noalign{\vskip 0.5ex}
& \textbf{\proposed~(Ours)} & \textbf{91.33} & \textbf{80.56} & \textbf{66.15} & \textbf{60.02} & \textbf{78.44} & \textbf{75.30} \\
\bottomrule
\end{tabular}
}
\end{table*}

\begin{figure*}
    \centering
    \subfloat[SVHN]{
        \includegraphics[width=0.3\linewidth]{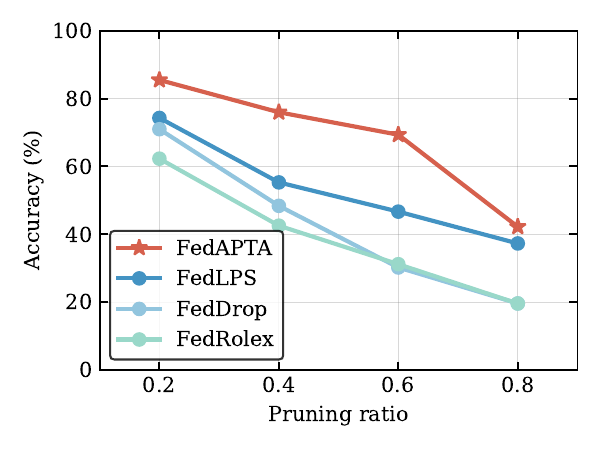}
        \label{fig:linechart:svhn}
    }\hfill
    \subfloat[CIFAR10]{
        \includegraphics[width=0.3\linewidth]{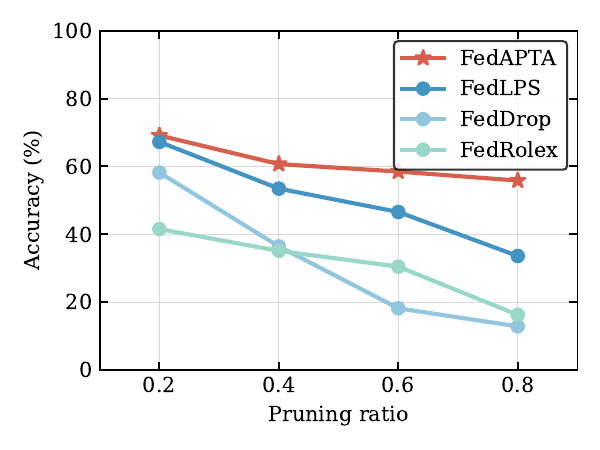}
        \label{fig:linechart:cifar10}
    }\hfill
    \subfloat[EMNIST]{
        \includegraphics[width=0.3\linewidth]{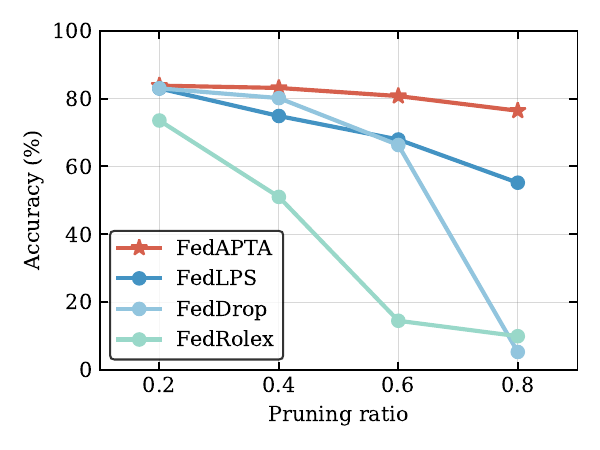}
        \label{fig:linechart:emnist}
    }
    \caption{Model accuracy of FedAPTA, FedLPS~\cite{jia2024fedlps}, FedDrop~\cite{caldas2018expanding_feddrop}, and FedRolex~\cite{alam2022fedrolex} on the RseNet18 model with different pruning ratios on the non-i.i.d. setting of the SVHN, CIFAR10, and EMNIST datasets.}
    \label{fig:linechart}
\end{figure*}

\begin{table}
\centering
\caption{Number of parameters of ResNet18 and ShuffleNetV2 under different pruning ratios in \proposed.}
\renewcommand{\arraystretch}{1.2}
\begin{tabular}{ccc}
\toprule
\textbf{Pruning ratios} 
& \textbf{ResNet18} & \textbf{ShuffleNetV2} \\
\midrule
0   & 11.01M & 1.21M  \\
0.2 & 8.81M  & 0.97M  \\
0.4 & 6.61M  & 0.72M  \\
0.6 & 4.40M  & 0.48M  \\
0.8 & 2.20M  & 0.24M  \\
\bottomrule
\end{tabular}
\label{tab:params}
\end{table}

\subsection{Effect of Pruning Ratios}
In FL frameworks where model pruning is used, the number of parameters in the model to be trained is significantly reduced, thereby decreasing the computing resources required for training local models, as demonstrated in \Cref{tab:params}. However, as the pruning ratio increases, the model's accuracy tends to deteriorate. Therefore, in this subsection, we investigate how varying pruning ratios affect model performance. In this experiment, we compare the proposed adaptive layer-wise model pruning method to the pruning methods used in FedLPS~\cite{jia2024fedlps}, FedDrop~\cite{caldas2018expanding_feddrop}, and FedRolex~\cite{alam2022fedrolex}, under the setting of $\rho=\{0.2,0.4,0.6,0.8\}$. \Cref{fig:linechart} illustrates the model accuracy of the ResNet-18 model after 100 communication rounds of FL under non-i.i.d. settings on the SVHN, CIFAR-10, and EMNIST datasets. The results indicate that, on these three datasets, \proposed~consistently outperforms the aforementioned pruning methods when the pruning ratio $\rho$ takes values between 0.2 and 0.8, and is capable of maintaining satisfactory model accuracy even under relatively high pruning rates.

\subsection{Effects of Model Similarity Metrics}
\begin{figure}
    \centering
    \subfloat[Non-i.i.d.]{
        \includegraphics[width=0.7\linewidth]{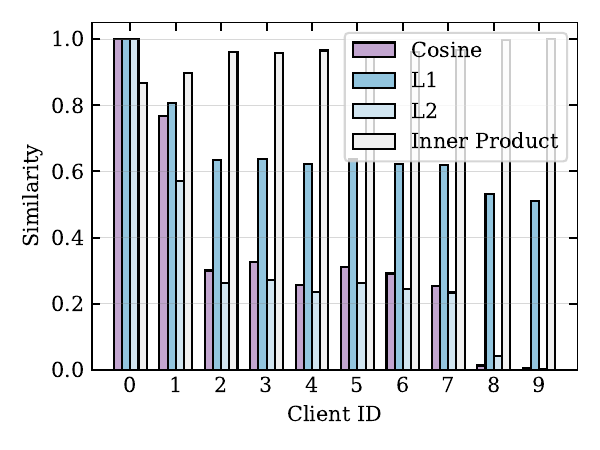}
        \label{fig:metrics:noniid}
    }\hfill
    \subfloat[i.i.d.]{
        \includegraphics[width=0.7\linewidth]{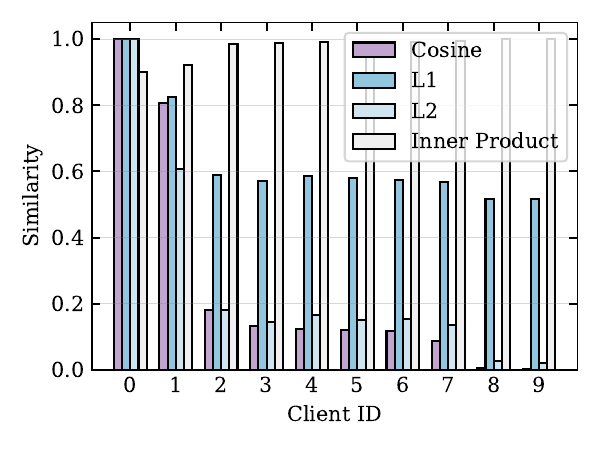}
        \label{fig:metrics:iid}
    }
    \caption{Effects of model similarity metrics, where device 0 and device 1 handle the same task. Cosine similarity best captures the relationship between devices 0 and 1 since it produces the most distinguishable similarity values among all devices.}
    \label{fig:metrics}
\end{figure}

A key design of \proposed~is that we use cosine similarity as the basis for clustering and perform the aggregation of local models based on the clustering results. In this context, we conduct a comparative analysis among four distinct types of model similarity metrics, including L1, L2, inner product, and cosine similarity. We conduct experiments on the ResNet-18 model, where device 0 and device 1 were configured to possess similar data distributions. Then, we measure the similarity between the local model of device 0 and all other devices. It should be emphasized that, for the purpose of achieving a clearer and more interpretable comparison, the similarity values were normalized based on the self-similarity values of device 0. The results under both non-i.i.d. and i.i.d. settings are presented in \Cref{fig:metrics}, respectively. As illustrated in the figure, cosine similarity demonstrates the most effective capability in capturing the similarity between device 0 and device 1, since it produces the most distinguishable similarity values among all devices.

\begin{table*}
\centering
\renewcommand{\arraystretch}{1.2}
\caption{Existing work in FL.}
\label{tab:related works}
\resizebox{\linewidth}{!}
{
\begin{tabular}{lllccc}
\toprule
\textbf{Category} & \textbf{Frameworks} & \textbf{Main Techniques} & \textbf{Multi-task} & \textbf{Model Pruning} \\
\midrule
& FedDrop~\cite{caldas2018expanding_feddrop} & Randomly drop to obtain a sub-model & $\times$ & $\checkmark$ \\
& FedRolex~\cite{alam2022fedrolex} & Uses a rolling window to extract parts of a global model & $\times$ & $\checkmark$ \\
FL with pruning & Hermes~\cite{li2021hermes} & Applies structured pruning to obtain a sub-model & $\times$ & $\checkmark$ \\
& FedMP~\cite{jiang2022fedmp} & Dynamic pruning ratios via multi-armed bandit algorithm & $\times$ & $\checkmark$ \\
& FedLPS~\cite{jia2024fedlps} & Adaptive channel-wise pruning & $\checkmark$ & $\checkmark$ \\
[0.3ex]\hline\noalign{\vskip 0.5ex}
& FedEM~\cite{marfoq2021federated} & Hybrid distribution modeling & $\checkmark$ & $\times$ \\
& MTFL~\cite{mills2021multi} & Allows devices to train personalized models & $\checkmark$ & $\times$ \\
FL in multi-task & FedMSplit~\cite{chen2022fedmsplit} & Dynamic graphs for relationships in multimodal models & $\checkmark$ & $\times$ \\
scenarios & MOCHA~\cite{smith2017federated_mocha} & Modeling inter-device relationships & $\checkmark$ & $\times$ \\
& Ghosh et al.~\cite{ghosh2019robust_gho} & Group devices by data distribution through robust clustering & $\checkmark$ & $\times$ \\
& CFL~\cite{sattler2020clustered_cfl} & Device clustering based on gradients & $\checkmark$ & $\times$ \\
& ClusterFL~\cite{ouyang2021clusterfl} & Uses a cluster indicator matrix Selects relevant devices & $\checkmark$ & $\times$ \\
& FedDGA~\cite{10382491} & Dynamic guided attention & $\checkmark$ & $\times$ \\
\bottomrule
\end{tabular}
}
\end{table*}

\section{Related Work}

    \subsection{Federated Learning}
    FL is a distributed ML framework that is explicitly designed to enable the collaborative training of a global model across multiple devices, while simultaneously preserving and safeguarding data privacy~\cite{mcmahan2017communicationfedavg}. Within the paradigm of FL, each participating device utilizes its locally stored data to perform a number of local training steps, after that, the updated model parameters are transmitted to a central server. The central server subsequently performs an aggregation operation over all the received parameters from the various devices, thereby updating the shared global model. Importantly, throughout this entire process, only the updated models are communicated to the central server, and there is no need to transmit any raw local data.
    
    However, when deploying deep neural networks (DNNs) in the FL framework, the system often incurs substantial communication and computation overhead~\cite{huang2017densely_dnn}. Various efforts have attempted to alleviate this issue through a number of techniques, such as reducing the frequency of model aggregation on the central server~\cite{mcmahan2017communicationfedavg}, applying parameter sparsification~\cite{han2020adaptive_dense1, hu2021federated_dense2_new}, and quantization~\cite{sattler2019robust_liang1, hubara2018quantized_liang2} techniques. However, these approaches have predominantly focused on improving communication efficiency, while overlooking the fact that devices often exhibit limited and heterogeneous computational capabilities~\cite{gao2022survey_hetero}. This heterogeneity among devices typically leads to performance degradation in real-world FL. This paper focuses on the impact of this computational capability heterogeneity on FL.
    
    \subsection{Model Pruning in Federated Learning}
    With the continuous advancement of DNNs, the increasing number of parameters and the resulting computational overhead have significantly slowed down model training. To tackle this pressing challenge, a pruning method has been introduced by \cite{han2015learning_pruningst}, aiming to compress DNNs by removing redundant parameters and structural components. This approach seeks to reduce both computational and storage costs, while maintaining model performance~\cite{liu2017learning_pr1, he2017channel_pr2, han2015learning_prCompressParams, lin2018accelerating_prStr}.

    In FL, the large size of models has emerged as a significant bottleneck, impeding their efficient deployment and training on resource-constrained devices~\cite{jiang2022fedmp}. 
    Since the central server cannot directly access local data, traditional pruning strategies~\cite{han2015learning_pruningst,fang2023depgraph_pruning} that depend on global data distribution are difficult to apply directly in FL settings. To address this limitation, prior research has investigated incorporating model pruning into FL frameworks as a means to reduce the number of model parameters. For example, FedDrop~\cite{caldas2018expanding_feddrop} trains only a subset of the global model on each device to lower local computation costs. FedRolex~\cite{alam2022fedrolex} utilizes a sliding window mechanism to extract sub-models, enabling balanced training across different parts of the global model; Hermes~\cite{li2021hermes} applies structured pruning on devices to obtain lightweight sub-models. FedMP~\cite{jiang2022fedmp} employs a multi-armed bandit algorithm to dynamically adjust the pruning ratio per device. FedLPS~\cite{jia2024fedlps} introduces an adaptive channel-wise pruning algorithm to produce task-specific lightweight predictors. However, most existing approaches adopt a uniform pruning ratio across all candidate layers, without considering the varying importance of individual layers within the model.

    \subsection{Federated Multi-Task Learning}
    In practical FL applications, it is common for devices to perform different tasks. Existing studies, such as Hermes~\cite{li2021hermes}, have shown that in such multi-task learning scenarios, the performance of a globally trained model may fall short of that achieved by models trained locally, which fundamentally contradicts the core objective of FL.

To address this issue, several works have explored the development of personalized models. For instance, FedEM~\cite{marfoq2021federated} models each device’s data distribution as a mixture of multiple latent components, enabling the joint learning of shared component models and personalized mixture weights, which significantly enhances both model accuracy and fairness. MTFL~\cite{mills2021multi} introduces non-federated batch normalization layers, allowing devices to train personalized deep neural networks, thereby improving accuracy and convergence speed. FedMSplit~\cite{chen2022fedmsplit} employs a dynamic graph structure to represent relationships among multimodal device models, enabling the learning of personalized yet globally correlated local models.
However, it is important to note that these approaches primarily focus on training models tailored to individual devices. 

A few studies have further explored multi-task learning in the context of FL. For example, MOCHA~\cite{smith2017federated_mocha} applies a multi-task learning framework to train multiple models across devices. Ghosh et al.\cite{ghosh2019robust_gho} propose a robust clustering algorithm that groups devices with similar data distributions while resisting interference from Byzantine devices. CFL\cite{sattler2020clustered_cfl} clusters devices based on the similarity of their gradient updates, enabling the learning of specialized models within each cluster—particularly beneficial in non-i.i.d. settings. ClusterFL~\cite{ouyang2021clusterfl} enhances training efficiency by discarding straggling devices and selecting those with greater relevance through a cluster indicator matrix.
FedDGA \cite{10382491} addresses non-i.i.d. data distribution and task imbalance issues by introducing Dynamic Guided Attention (DGA) for adaptive feature alignment and Dynamic Batch Weighting (DBW) for loss balancing, achieving a high accuracy while maintaining communication efficiency. 

\section{Conclusion}
In this paper, we propose \proposed, 
a FL framework designed to address the challenge of heterogeneous computational capabilities and the suboptimal global model performance caused by directly aggregating local models trained on different tasks.
To mitigate the first issue, we design an adaptive layer-wise model pruning method in \proposed, which permits heterogeneous devices to participate in federated training with pruned models. Furthermore, \proposed~introduces a novel heterogeneous model recovery algorithm, which is capable of effectively aggregating pruned models with the assistance of the latest global model information. 
To alleviate the second issue, \proposed~achieves task-aware model aggregation by performing device clustering. Specifically, it enables local models to be aggregated separately for each task. 
Extensive experimental comparisons demonstrate that \proposed~is capable of preserving superior model accuracy across a range of FL scenarios.
In this future, we plan to explore \proposed's~performance in other task domains beyond classification.

\section*{Acknowledgments}
This work was supported in part by the National Natural Science Foundation of China (62406215 and 62072321), the Science and Technology Program of Jiangsu Province (BZ2024062), the Natural Science Foundation of the Jiangsu Higher Education Institutions of China  (22KJA520007), Suzhou Planning Project of Science and Technology (2023ss03).

\bibliographystyle{elsarticle-num} 
\bibliography{main}







\end{document}